\documentclass[12pt]{article}
\usepackage[a4paper]{geometry}
\usepackage{url}
  \let\oldurl\url
\usepackage{hyperref}
  \let\linkurl\url
  \let\url\oldurl
\usepackage{cleveref}
\usepackage{wrapfig}
\usepackage[english]{babel}
\usepackage[utf8x]{inputenc}
\usepackage[T1]{fontenc}
\usepackage{times}
\usepackage{amsmath}
\usepackage{color}
\usepackage{amssymb}
\usepackage{graphicx}
\usepackage{caption2}
\usepackage{hang}
\usepackage{harvard}
\usepackage{algorithm,algpseudocode}
\usepackage{amsmath}
\usepackage{graphicx}
\usepackage[T1]{fontenc}
\usepackage{array}
\usepackage{makecell}
\newcolumntype{x}[1]{>{\centering\arraybackslash}p{#1}}

\usepackage{tikz}

\newtheorem{task}{Task}[section]
\newtheorem{definition}{Definition}[section]

\title{Application of Bayesian Networks for Estimation of Individual Psychological Characteristics}
\author{Alexander LITVINENKO$^1$, Natalya LITVINENKO$^2$, Orken MAMYRBAYEV$^3$}

\begin{document}
\maketitle
%\myaddress
\begin{center}
$^1$Extreme Computing Research Center, Center for Uncertainty Quantification in Computational Science \& Engineering King Abdullah University of Science and Technology, Thuwal 23955-6900, Kingdom of Saudi Arabia, e-mail: alexander.litvinenko@kaust.edu.sa,\\
$^2$Applied research department, Institute of Mathematics and Mathematical Modeling CS MES RK, Almaty, Kazakhstan, e-mail: n.litvinenko@inbox.ru\\
$^3$Institute of information and computational technologies CS MES RK, Almaty, Kazakhstan,
e-mail: morkenj@mail.ru
\end{center}

\begin{abstract}
An accurate qualitative and comprehensive assessment of human potential is one of the most important challenges in any company or collective. We apply Bayesian networks for developing more accurate overall estimations of psychological characteristics of an individual, based on psychological test results, which identify how much an individual possesses a certain trait. Examples of traits could be a stress resistance, the readiness to take a risk, the ability to concentrate on certain complicated work. The most common way of studying psychological characteristics of each individual is testing. Additionally, the overall estimation is usually based on personal experiences and the subjective perception of a psychologist or a group of psychologists about the investigated psychological personality traits.
 
\end{abstract}
%\mykeywords

\section{Introduction}
In this article we discuss applications of Bayesian network methods for solving typical and highly demanding tasks in psychology. 
We compute overall estimates of the psychological personality traits, based on given answers on offered psychological tests, as well as a comprehensive study of the social status of the individual, their religious beliefs, educational level, intellectual capabilities, the influence of a particular social environment, etc.
We believe that the most optimal mathematical model for solving this problem is a graphical probabilistic model with strongly expressed cause-effect relations. Therefore, we chose the Bayesian network as our model. Advantages of the Bayesian network are as follows: 1) The Bayesian network reflects the causal-effect relationship very well. 2) The mathematical apparatus of Bayesian networks is well developed and thus, there are many software implementations of the Bayesian network methods available.

Bayesian network is a graphical probabilistic model that represents a set of random variables and their conditional dependencies via a directed acyclic graph \cite{BenGal08},
\cite{Pourret08}, \cite{Albert:09}. 
For example, a Bayesian network could represent the probabilistic connections between overall economical situations, average salaries and nationalism in society. It can give recommendations to local governments of which steps to undertake to decrease the level of political tensions. Other promising applications are in Human Resource (HR) departments and in marriage agencies. Bayesian networks, by analyzing psychological properties of each individual, and sociological connections between individuals, may help to select a better group for a certain task, prevent possible conflicts and increase performance.

Bayesian framework is very popular in various kinds of applications: parameter identification \cite{matthies2016parameter}; Bayesian update  \cite{matthies2016bayesian}, \cite{Rosic2013}; uncertainty quantification \cite{rosic2012sampling}, \cite{rosic2011direct}, \cite{pajonk2012deterministic}, \cite{UQLitvinenko12}; inverse problems \cite{hermann2016inverse}; classification \cite{berikov2003influence}, \cite{berikov2003methods}, \cite{berikov2004discrete}.

In this work we will apply Bayesian network \cite{Albert:09} to find a more accurate overall estimate for each investigated psychological personality trait (PPT), see Definition \ref{definition3}. Our mathematical model for the construction of overall estimate is the graphical probabilistic model that reflects probabilistic dependencies between the questions used in psychological tests and the overall estimates of the investigated PPT. Due to the presence of cause-effect relationships we will use Bayesian networks as the graphical probabilistic model \cite{Tu:06}. We consider also some of the problems which can typically arise during the computing of the overall estimates. For these problems we describe a step-by-step construction of the Bayesian network and we provide the programming code.
In the world of psychological tests, there are special software products that help specialists develop new tests and adapt existing products.

The main goals of this work are as follows:
\begin{enumerate}
\item to develop principles for constructing the overall estimates of PPT based on the usage of the Bayesian network;
\item to demonstrate the potential of graphical probabilistic models for solving problems of this type on several easy examples;
\item to implement these examples in R programming language;
\item to show the capabilities of Bayesian network for a qualitative analysis of the obtained solution.
\end{enumerate}
The structure of the paper is as follows: In Section~\ref{sec:defi} we introduce the required notions and definitions. Section~\ref{sec:ProblemStatement} is devoted to the problem statement. In Section~\ref{sec:ExamplesTasks} we consider and solve three different examples. We also list the solution in R-code. Finally, in the conclusion we repeat the main achievements and share our experience.%
\section{Notions and definitions}
\label{sec:defi}
In this section we list the necessary definitions that will be used below. These definitions do not always coincide with definitions used in similar works. There are different reasons for this:
\begin{itemize} 
\item many terms and definitions in psychology are not yet completely formed;
\item the meaning of the proposed concepts does not contradict the common notions in other literature;
\item our definitions simplify presentation and reading.
\end{itemize} 
\begin{definition}
\textbf{Latency} is the property of objects or processes to be in a hidden state, without demonstrating itself explicitly.
 \label{definition1}
\end{definition}
\begin{definition}
\textbf{Psychological test} is a standardized exercise, which results provide information about certain psychological traits of the individual
 \label{definition2}
\end{definition}
\begin{definition}
\textbf{Psychological trait} is any stable characteristic of a person. This characteristic can be spiritual, moral, social, and is the real reason for the specific behavior of a given person under certain conditions.
 \label{definition3}
\end{definition}
\begin{definition}
\textbf{A priori estimate} is the estimate, obtained before the experiment, on the basis of expert knowledge, some additional information, or taken as a first approximation.
 \label{definition4}
\end{definition}
\begin{definition}
\textbf{A posteriori estimate} is the estimate obtained after the experiment, based on the results of this experiment.
 \label{definition5}
\end{definition}
\begin{definition}
\textbf{Graph} is a set of vertices (nodes) connected by edges. We can also say that graph G is a pair of sets $G = (V,E)$, where $V$ is a subset of any countable set, and $E$ is a subset of $V \times V$. An oriented graph is a graph with oriented edges.
 \label{definition6}
\end{definition}
\begin{definition}
\textbf{Graphical probabilistic model} is a probabilistic model, where the graph shows the probabilistic relationship between random variables. The vertices of the graph are random variables, and the edges reflect the probabilistic relationships between random variables.
 \label{definition7}
\end{definition}
In the current work the vertices reflect investigated traits and estimates, and the edges reflect dependencies between traits and estimates.

\begin{definition}
\textbf{Bayesian network} is the finite, oriented and acyclic graph representing the graphical probability model.
 \label{definition8}
\end{definition}
\begin{table}[!ht]
\begin{center}
\begin{tabular}{|c|l|}
\hline
\textit {Notation} & Meaning  \\
\hline
PPT & psychological personality trait(s)  \\
\hline
$N$ & number of psychological traits  \\
\hline
$F_j$ & investigated psychological traits, $j=1,2,...,N$  \\
\hline
$E_j$ & a level how a respondent possesses trait $F_j$, $j=1,2,...,N$  \\
\hline
$Q_{jk}$ & a question, $j=1,2,...,N$,  $k=1,2,...,M_j$.\\

&There is a set of questions $\{Q_{jk}\}$ for each investigated trait $F_j$\\
\hline
$E_{jk}$ & a grade, which a respondent received for his answer on question $Q_{jk}$\\
\hline
\end{tabular}
\caption{Notation}
\end{center}
\end{table}
\section{Problem statement}
\label{sec:ProblemStatement}
Let us construct a psychological test. In this test we investigate traits $F_j$, $j=1,2,...,N$. Here $F_1$ could be the stress resistance of an individual, $F_2$ the ability to speak the Spanish language, and so on. Suppose that the current respondent possesses all these traits $F_j$. The level how this respondent possesses trait $F_j$ will be denoted by $E_j$. For instance, $E_j=5$ (for the scale 1-very bad, 2-bad, 3-satisfactory, 4-good, 5 excellent) means that the respondent speaks excellent Spanish. Note that $F_j$ are latent values, i.e., for instance, it is hard to say without testing if the respondent possesses stress resistance or not. These $E_j$ estimates depend on many subjective items such as the quality of the tests, the mental state of the respondent, the psychologist's perceptions, etc. 

For each investigated trait $F_j$ there is a set of questions $\{Q_{jk}\}$,  $j=1,2,...,N$,  $k=1,2,...,M_j$ in the test. A respondent receives a grade $E_{jk}$ for his answer on question $Q_{jk}$, where index $j$ means the $j$-th trait and index $k$ the $k$-th question.

A professional psychologist can conduct the dependence between the received grades $E_{jk}$ and the presence of $F_j$ by the given respondent. Such type of a priori data can also be obtained from previous experimental data, theoretical knowledge or psychologist's perception. Later on we will start with a Bayesian network that contains a priori probabilities in each node. Then our algorithm will visit each node and recompute each probability value inside (compute posterior probabilities). 
 
The scales of grades for $E_j$ and $E_{jk}$ are discrete and can be very different. Often, scales are chosen on the basis of the subjective wishes of the psychologist. Note that scales for different values may not coincide. The most common scales are:
\begin{itemize}	
\item a two-point scale \{0,1\}, \{Yes, No\}, \{presence, absence\}, \{true, false\};
\item a three-point scale \{-1, 0, 1\}, or, for example, \{coward, balanced, brave\}, \{absence, limited presence, presence\}, \{0. 1, 2\}, etc.
\item a five-point scale \{1, 2, 3, 4, 5\} or \{0, 1, 2, 3, 4\} etc. 
\item a ten-point scale \{1, 2, ..., 10\}. 
\item a hundred-point scale \{1, 2, ..., 100\}. 
\end{itemize}

The final aim is to assess the presence of PPT by the respondent, based on the a priori estimates of the psychologist and answers of the respondent on the testing questions. It is necessary to take into account the following points:
\begin{itemize}	
\item there is the probability of a ``slip'', i. e., when the respondent occasionally gives a wrong answer (pressed a wrong button on the keyboard). 
\item there is a certain chance of occasionally guessing the correct answer.
\end{itemize}
\section{Examples of tests}
\label{sec:ExamplesTasks}
In this section we consider three typical test examples with increasing complexity. All three examples include tables with a priori estimates, given in a table format, as well as a description of these tables.
After each description we formulate three possible quantities of interests (we call these quantities of interests - ``Tasks''). After that we demonstrate how these Tasks can be solved with the Bayesian network.
 
The first test example is simple, the posterior is equal to prior. In the second and third test examples we will consider three Tasks. Each example contains the formulation of quantity which should be computed, settings of all required parameters and the corresponding solution in the R-code. 

\subsection{An example of a test with one question}
\label{sec:Ex1}
We consider the simplest situation with one question and one trait (PPT). In this case, the simplest graph consists of two vertices and one edge -- one vertex for the question and one vertex for the trait and the edge connect these vertices.

\begin{figure}[!ht]
 \centering
  \includegraphics[width=0.5\textwidth]{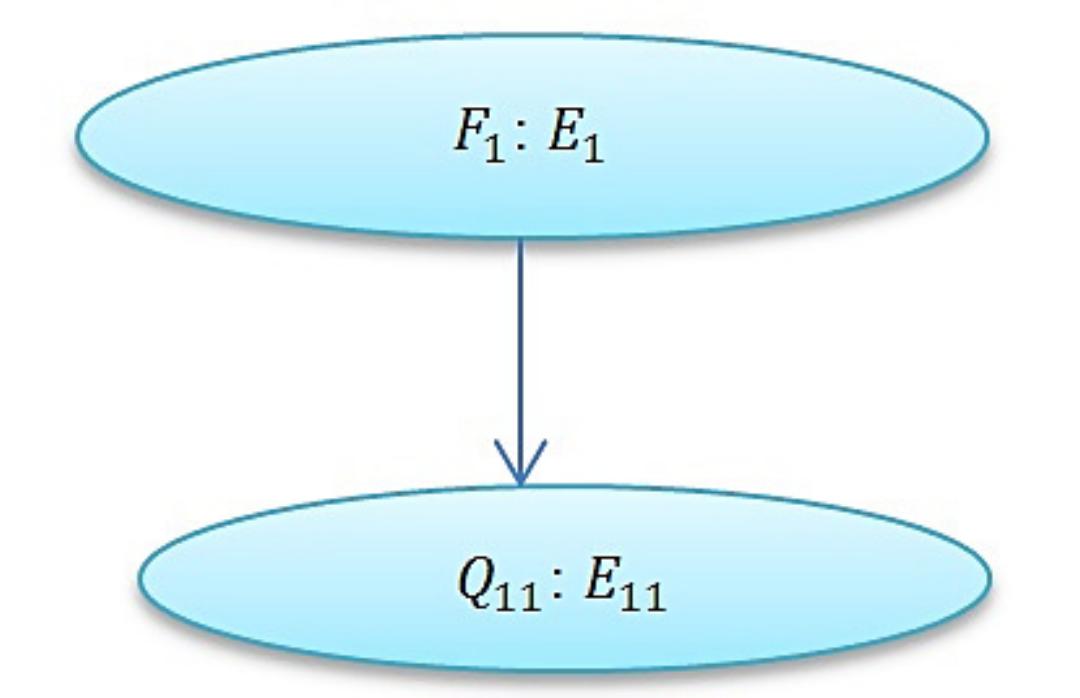}
 \caption{Bayesian network, where the estimate of the trait $F_1$ is determined by one question $Q_{11}$.}
  \label{fig:fig1}
\end{figure}

Let us compute the grade $E_1$ how the respondent possesses the trait $F_1$. 
For this PPT $F_1$ there is only one question $Q_{11}$ in this test. We denote the grade, received by the respondent by answering on this question as $E_{11}$.
A respondent may possess this PPT with varying degrees. Depending on the primary (a priori) grade of the trait $F_1$, one can a priori assume how respondents will answer question $Q_{11}$.  Corresponding Bayesian network is in Fig.~\ref{fig:fig1}.

\begin{figure}[!ht]
\centering
\includegraphics[width=0.95\linewidth]{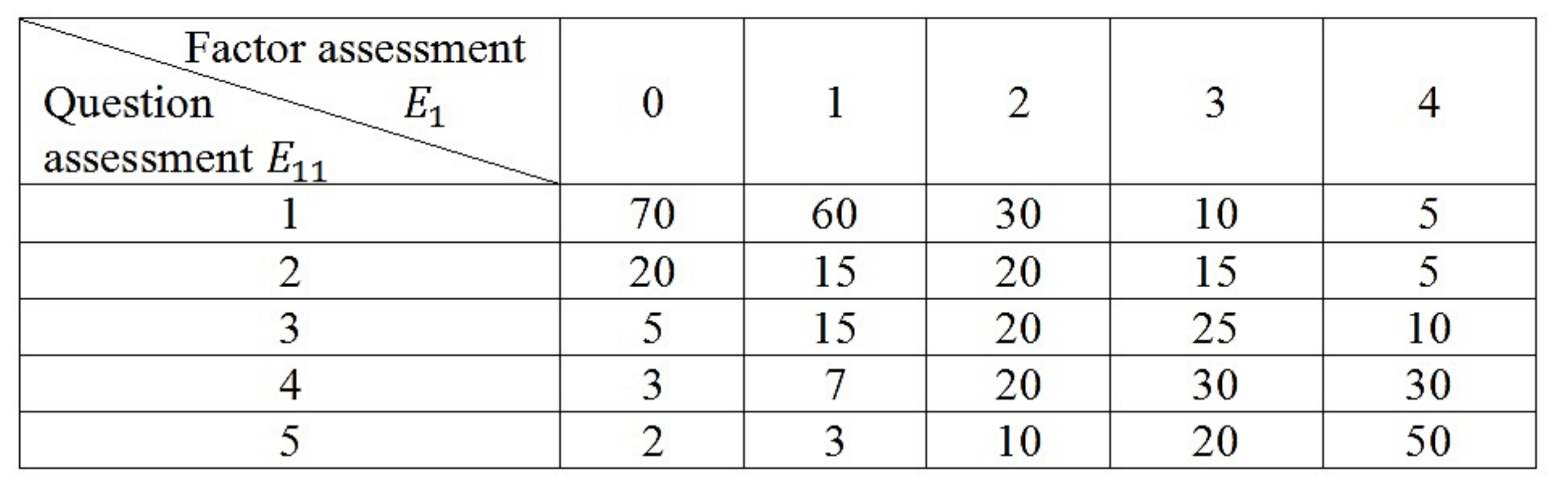}
\captionof{table}[foo]{A priori estimates for the test example from Section~\ref {sec:Ex1}, where the overall grade of the trait $F_1$ is determined by one question $Q_{11}$.}
  \label{tab1}
\end{figure}

In this example the overall grade is the same as for the single test question for this trait. There is nothing to compute here. A psychologist usually has some a priori knowledge of how a respondent, who possess a certain trait, can respond to a single test-question for this trait. A priori knowledge is usually obtained from previous experiments, or relies on the knowledge of a psychologist. Table~\ref{tab1} gives prior estimates for this example. The trait is estimated by a discrete number from the set $\{1, 2, 3, 4, 5\}$. The question is estimated also by a discrete value from the set $\{1, 2, 3, 4, 5\}$.

One can interpret values from Table~\ref{tab1} in the following way.  A psychologist thinks that:
\begin{enumerate}
\item 70\% of respondents, who do not possess the trait $F_1$ 
($E_1 = 0$) will answer question $Q_{11}$ with grade $E_{11}=1$.
\item only 20\% of respondents, who possess the trait $F_1$ with grade ($E_1 = 3$) will answer question $Q_{11}$ with grade $E_{11}=5$.
\item only 2\% of respondents, who do not possess the trait $F_1$ ($E_1 = 0$) will answer question $Q_{11}$ with grade $E_{11}=5$.
\end{enumerate}	

This example is simple, the posterior probabilities are equal to prior probabilities. Table~\ref{tab1} gives us all necessary information. The situation becomes more complicated if there are several questions.
\subsection{An example of a test with two questions}
\label{sec:Ex2}
\begin{figure}[!ht]
 \centering
  \includegraphics[width=0.95\textwidth]{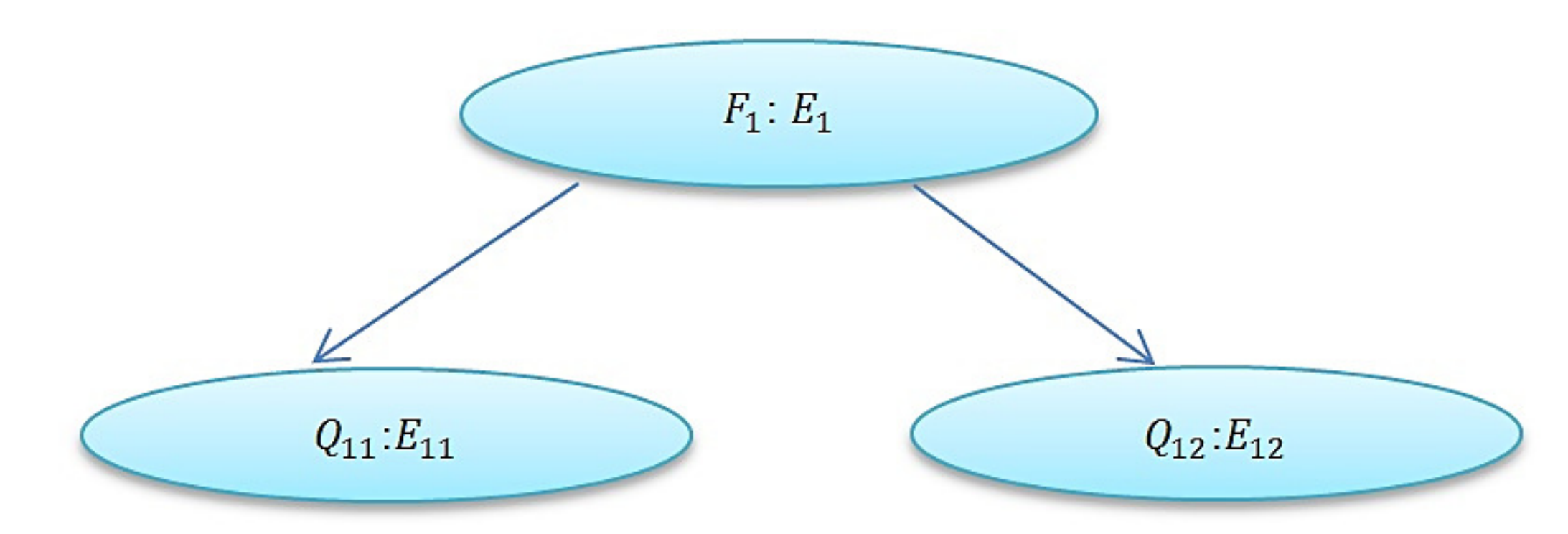}
 \caption{Bayesian network, where the estimate of the trait $F_1$ is determined by two questions $Q_{11}$ and $Q_{12}$.}
  \label{fig:fig2}
\end{figure}

We increase the complexity of the example from Section~\ref{sec:Ex1}, namely, we consider two questions in the test. The PPT is estimated on a two-point scale. The question is estimated on a five-point scale. Therefore, the graph consists of 3 vertices and 2 edges \cite{Mu:06} (Fig.~\ref{fig:fig2}).

\begin{figure}[!ht]
\centering
\includegraphics[width=0.95\linewidth]{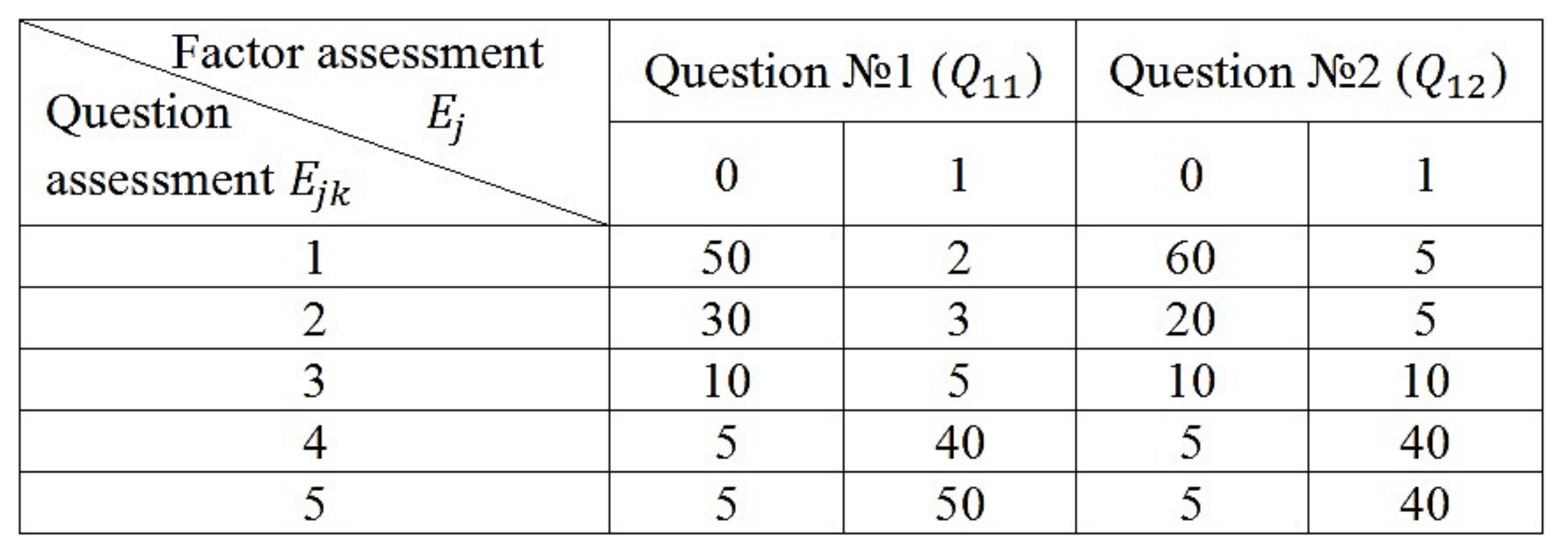}
\captionof{table}[foo]{Prior estimates for the example from Section~\ref{sec:Ex2}, where the estimate of the trait $F_1$ is determined by two questions $Q_{11}$ and $Q_{12}$.}
 \label{tab2}
\end{figure}

Let us build the estimate $E_1$ of $F_1$. For this trait there are two questions $Q_{11}$ and $Q_{12}$ in the test. We denote these estimates of questions by $E_{11}$ and $E_{12}$ respectively. Table~\ref{tab2} gives prior estimates for this example.

One can interpret the values from Table~\ref{tab2} in the following way.  A psychologist thinks that:
\begin{enumerate}
\item 60\%
of respondents, who do not possess the trait $F_1$ will answer question $Q_{12}$ with grade $E_{12}=1$.
\item 50\% of respondents, who possess the trait $F_1$ will answer question $Q_{11}$ with grade $E_{11}=5$. Also 40\% of respondents, who possess the trait $F_1$ will answer question $Q_{12}$ with grade $E_{12}=5$. 
\item 3\% of respondents, who do not possess the trait $F_1$ will answer question $Q_{11}$ with grade $E_{11}=5$. Also 5\% of respondents, who do not possess the trait $F_1$ will answer question $Q_{12}$ with grade $E_{12}=5$.
\end{enumerate}	
	
Possible quantities of interest in this example could be: 
\begin{enumerate}
\item What is the probability of receiving the grade $E_j$ for each question by a respondent if a priori probabilities are given as in Table~\ref{tab2}.  
\item The respondent answered the first question with grade 2, the second question with grade 3. What is the probability that the respondent possesses the trait $F_1$?
\item The respondent answered the first question with grade 3. What is the probability that the respondent will answer the second question with grades 4 or 5?
\end{enumerate}	
To compute these three possible quantities of interest in R environment, we run the following commands as in Algorithm~\ref{a:alg1}. This preprocessing code allow us to include required R packages.
\begin{algorithm}
\begin{small}
\caption{R settings}
\label{a:alg1}
\begin{algorithmic}
\State \text{\#Clear the screen}
\State \text{rm(list=ls(all=TRUE))}
\State \text{\#Call the library bioconductor}
\State \text{source(``http://bioconductor.org/biocLite.R")}
\State \text{biocLite(``RBGL")}
\State \text{biocLite(``Rgraphviz")}
\State \text{\#Set all libraries we need}
\State \text{install.packages(``gRbase")}
\State \text{install.packages(``gRain")}
\State \text{library(gRbase)}
\State \text{library(gRain)}
\State \text{library(Rgraphviz)}
\end{algorithmic}
\end{small}
\end{algorithm}
Now we list the required steps in R environment, which set a priori distributions and build preliminary Bayesian network for all three Tasks.

\begin{algorithm}[!ht]
\caption{A priori parameter settings for the Example from Section~\ref{sec:Ex2}}
\begin{small}
\begin{algorithmic}
\label{a:alg2}
\State \text{\#Set a two-point scale for the given trait}
\State \text{lvl $\leftarrow$ c(``0",``1")}
\State \text{\#Set a five-point scale for questions}
\State \text{marks }$\gets$ \text{c(``1",``2",``3",``4",``5")}
\State \text{\#Assume a prior probability that the respondent possesses the given trait is 50\% }
\State \text{$F$ $\leftarrow$ cptable($\sim F$, values=c(50,50), levels=lvl)}
\State \text{\#Set a priori probabilities}
\State \text{$Q_{11}.F$ $\leftarrow$ cptable($\sim Q_{11} \mid F$,values=c(50, 30, 10, 5, 5,  2, 3, 5, 40, 50), levels=marks)}
\State \text{$Q_{12}.F$ $\leftarrow$ cptable($\sim Q_{12} \mid F$,values=c(60, 20, 10, 5, 5,  5,}
\text{5, 10, 40, 40), levels=marks)}
\State \text{\#Plot the graph}
\State \text{cpt.list $\leftarrow$ compileCPT(list($F$, $Q_{11}.F$, $Q_{12}.F$))}
\State \text{bnet $\leftarrow$ grain(cpt.list)}
\State \text{bnet $\leftarrow$ compile(bnet)}
\State \text{plot (bnet\$dag)}
\end{algorithmic}
\end{small}
\end{algorithm}
Now we formulate the Task:
\begin{task}
\label{t:2quest5}
To compute probability that a random respondent without any a priori knowledge about trait $F$ will answer on 2 questions.
\end{task}
Corresponding R-code, which solves this Task:
\begin{verbatim}
> xq1 = querygrain(bnet, nodes=c("Q11", "Q12"))
> xq1
$Q11
Q11
1    2    3    4    5 
0.26 0.17 0.08 0.23 0.28 
$Q12
Q12
1    2    3    4    5 
0.33 0.13 0.1  0.23 0.23 
\end{verbatim}
\textbf{Result:} From this listing in the R environment, one can see that due to prior data (in Table~\ref{tab2}), a respondent will answer the first question with grade, for example, 5, with probability 28\%, and with grade 3 with probability 8\%. Additionally, the last row shows that the respondent will answer the second question with grade 5 with probability 23\%, and with grade 3 with probability 10\%.
One more task is formulated as follows:
\begin{task}
\label{t:2quest1}
Assume that a respondent answered the first question with grade 2, the second question with grade 3. What is the probability that the respondent possesses trait $F_1$ ?\\
\end{task}
Corresponding R-code, which solves this Task:
\begin{verbatim}
> bnet.ev <- setEvidence(bnet, nodes = c("Q11","Q12"), 
states = c("2","3"))
xq2 = querygrain(bnet.ev,nodes=c("F"))
> xq2 = querygrain(bnet.ev, nodes=c("F"))
> xq2
$F
F
0        1 
0.91     0.09 
\end{verbatim}
\textbf{Result:} From the last line in the R environment, one can see that the respondent does not possess the trait with probability 91\% and possesses the trait with probability 9\%.
One more task is formulated as follows:
\begin{task}
\label{a:alg5}
Assume that the respondent answered the first question with grade 2. What is the probability that respondent will answer the second question with grade 4 or 5? 
\end{task}
Corresponding R-code is:
\begin{verbatim}
> bnet.ev <- setEvidence(bnet, nodes = c("Q11"), states = c("3"))
> xq2 = querygrain(bnet.ev, nodes=c("Q12"))
> xq2
$Q12
Q12
1    2    3   4    5 
0.42 0.15 0.1 0.17 0.17 
\end{verbatim}
\textbf{Result:} The respondent will answer the second question with grade 4 or 5 with probability 17\%+17\%=34\%.

\subsection{An example of test with five questions}
\label{sec:Ex3}
\begin{figure}[!ht]
 \centering
  \includegraphics[width=0.99\textwidth]{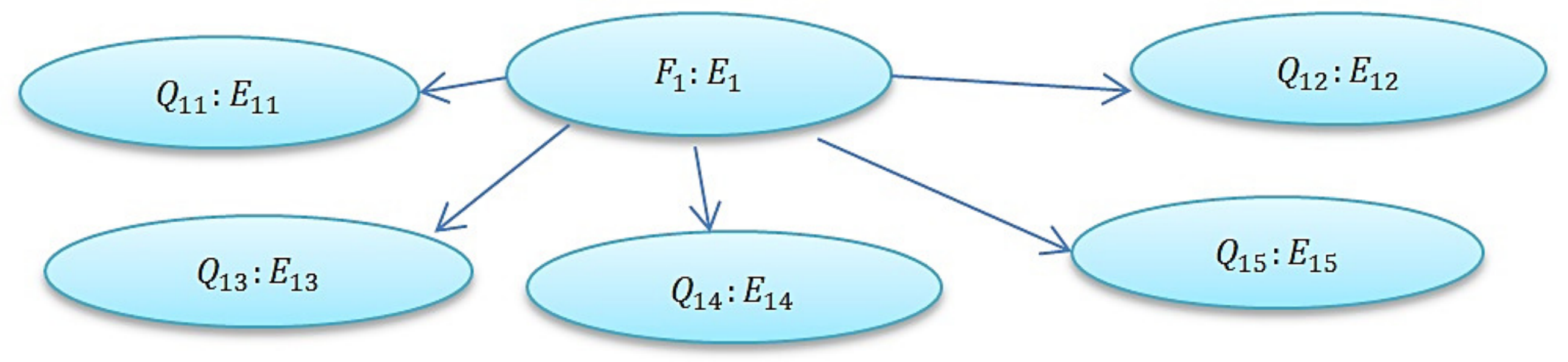}
 \caption{Bayesian network, where the estimate of the trait $F_1$ is determined by five questions.}
  \label{fig3}
\end{figure}

\begin{figure}[!ht]
\centering
\includegraphics[width=0.95\linewidth]{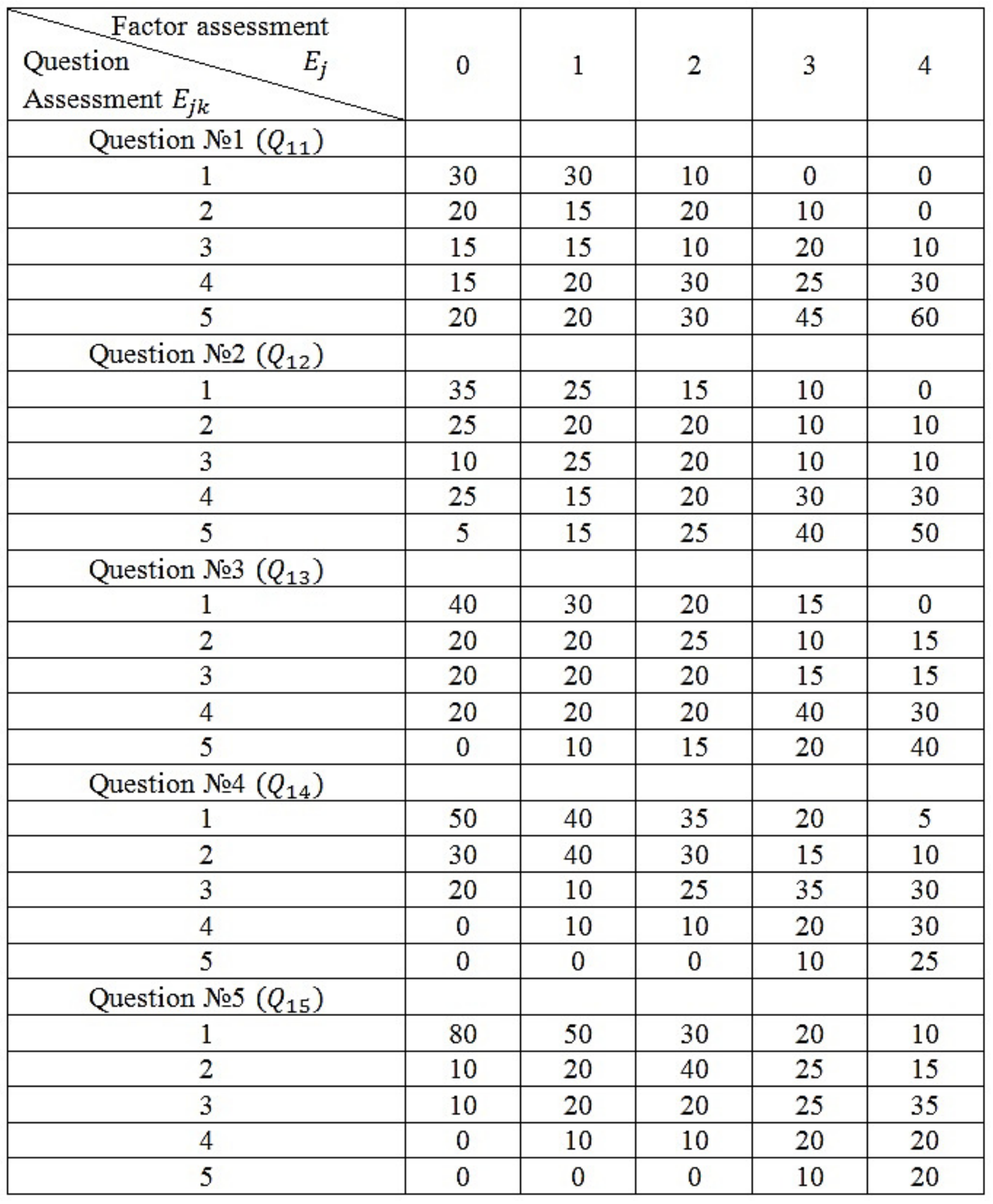}
\captionof{table}[foo]{A priori estimates for the test example from Section~\ref {sec:Ex3}, where the estimate of the trait $F_1$ is determined by five questions $Q_{11}$, $Q_{12}$,...,$Q_{15}$.
}
\label{tab3}
\end{figure}

In this example we will consider a test with 5 questions. For all 5 questions we set up a five-point scale. The corresponding graph (Fig.~\ref{fig3}) consists of 6 vertices (5 vertices for 5 questions and one vertex for PPT) and 5 edges (each edge connects a question with the trait ). 
Let us build the overall estimate $E_1$ for trait $F_1$. There are 5 questions $Q_{11}$, $Q_{12}$,..., $Q_{15}$ for this trait in the test. We denote estimates of these questions as $E_{11}$, $E_{12}$,...,$E_{15}$. 
Assume that experts, based on personal experience, have compiled Table~\ref{tab3} with prior estimates.

One can interpret the values from Table~\ref{tab3} in the following way.  A psychologist thinks that:
\begin{enumerate}
\item A priori it is known that respondents, who possess trait $F_1$ with grade $E_1=3$ will answer question $Q_{11}$ with grade $E_{11}=4$ (25\%), the question $Q_{12}$ with grade $E_{12}=4$ (30\%), the question $Q_{13}$ with grade $E_{13}=4$ (40\%), the question $Q_{14}$ with grade $E_{14}=4$  (20\%), the question $Q_{15}$ with grade $E_{15}=4$  (20\%).
\item A priori it is known that respondents, who possess trait $F_1$ with grade $E_1=0$ will answer question $Q_{11}$ with grade $E_{11}=4$ (15\%)
the question $Q_{12}$ with grade $E_{12}=4$ (25\%), the question $Q_{13}$ with grade $E_{13}=4$ (20\%), the question $Q_{14}$ with grade $E_{14}=4$ (0\%), the question $Q_{15}$ with grade $E_{15}=4$ (0\%). 
\item A priori it is known that respondents, who possess trait $F_1$ with grade $E_1=4$ will answer question $Q_{11}$ with grade $E_{11}=3$ (10\%), the question $Q_{12}$ with grade $E_{12}=3$ (10\%), the question $Q_{13}$ with grade $E_{13}=3$ (15\%), the question $Q_{14}$ with grade $E_{14}=3$ (30\%), the question $Q_{15}$ with grade $E_{15}=3$ (35\%). 
\end{enumerate}	

Possible quantities of interest here could be:
\begin{enumerate}
\item What is the probability that a respondent will answer all 5 questions with grade 5?
\item The respondent answered the first question with grade 5, the second and the third questions with grade 4. What is the probability that the respondent has the trait $F_1$ with grade not less than 3?
\item The respondent answered the first question with grade 5, the second and third questions with grade 3. What is the probability that the respondent will answer the fourth and fifth questions with grades not less than 4? 
\end{enumerate}	

The program code in R \cite{Ch:08}, computing quantities of interest, listed above, is the following. The setting commands for R are omitted for brevity. 
\begin{algorithm}[!ht]
\caption{A priory parameter settings for the Example from Section~\ref{sec:Ex3}}
\begin{small}
\begin{algorithmic}
\label{a:alg6}
\State \text{\#Set the five-point scale for the given trait}
\State \text{lvl $\leftarrow$ c("0","1","2","3","4")}
\State \text{\# Set the five-point scale for tests}
\State \text{marks $\leftarrow$ c("1","2","3","4","5")}
\State \text{\# A priori it is unknown if the respondent possesses the trait.} 
\State \text{\# With probability 20\% respondent can possess the trait in any amount.}
\State \text{$F$ $\leftarrow$ cptable($\sim F$, values=c(20,20,20,20,20),levels=lvl)}
\State \text{\#Set the a priori data (marks)}
\State \text{$Q_{11}.F$ $\leftarrow$ cptable($\sim Q_{11} \mid F$, values=c(30,20,15,15,}
\State \text{20,30,15,15,20,20,10,20,10,30,30,0,10,20,25, 45, 0, 0, 10, 30, 60), levels=marks)}
\State \text{$Q_{12}.F$ $\leftarrow$ cptable($\sim Q_{12} \mid F$, values=c(35,25,10,}
\State \text{ 25,5,25,20,25,15,15,15,20,20,20,25,10,10,10,30, 40, 0, 10, 10, 30, 50), levels=marks)}
\State \text{$Q_{13}.F$ $\leftarrow$ cptable($\sim Q_{13} \mid F$, values=c(40, 20, 20, 20,}
\State \text{0,  30, 20, 20, 20, 10, 20, 25, 20, 20, 15, 15, 10, 15, 40, 20, 0, 15, 15, 30, 40), levels=marks)}
\State \text{$Q_{14}.F$ $\leftarrow$ cptable($\sim Q_{14} \mid F$, values=c(50, 30, 20, 0, 0,}
\State \text{40, 40, 10, 10, 0, 35, 30, 25, 10, 0, 20, 15, 35, 20,10, 5, 10, 30, 30, 25), levels=marks)}
\State \text{$Q_{15}.F$ $\leftarrow$ cptable($\sim Q_{15} \mid F$, values=c(80, 10, 10, 0,}
\State \text{0,   50, 20, 20, 10, 0, 30, 40, 20, 10, 0, 20, 25, 25,20,10, 10, 15, 35, 20, 20), levels=marks)}
\State \text{\#Plot the graph}
\State \text{cpt.list $\leftarrow$ compileCPT(list($F, Q_{11}.F, Q_{12}.F,Q_{13}.F, Q_{14}.F, Q_{15}.F$))}
\State \text{bnet $\leftarrow$ grain(cpt.list)}
\State \text{bnet $\leftarrow$ compile(bnet)}
\State \text{plot (bnet\$dag)}
\end{algorithmic}
\end{small}
\end{algorithm}

\begin{task}
\label{t:2quest2}
Compute the probabilities that a respondent with no a priori information will answer all 5 questions.
\end{task}

Corresponding R-code, which solves this Task:
\begin{verbatim}
> xq1 = querygrain(bnet, nodes=c("Q11","Q12","Q13","Q14","Q15"))
> xq1
$Q11
Q11
1    2    3    4    5 
0.14 0.13 0.14 0.24 0.35 
$Q12
Q12
1    2    3    4    5 
0.17 0.17 0.15 0.24 0.27 
$Q13
Q13
1    2    3    4    5 
0.21 0.18 0.18 0.26 0.17 
$Q14
Q14
1    2    3    4    5 
0.30 0.25 0.24 0.14 0.07 
$Q15
Q15
1    2    3    4    5 
0.38 0.22 0.22 0.12 0.06
\end{verbatim}
The output of the R program can be interpreted as follows:
\begin{enumerate}
\item 
A random respondent will answer the first question with grade $\{1, 2, 3, 4, 5\}$ with probability \\$\{0.14, 0.13, 0.14, 0.24, 0.35\}$ respectively.
\item 
A random respondent will answer the second question  with grade $\{1, 2, 3, 4, 5\}$ with probability \\$\{0.17, 0.17, 0.15, 0.24, 0.27\}$ respectively.
\item 
A random respondent will answer the third question with grade $\{1, 2, 3, 4, 5\}$ with probability \\$\{0.21, 0.18, 0.18, 0.26, 0.17\}$ respectively.
\item 
A random respondent will answer the fourth question with grade $\{1, 2, 3, 4, 5\}$ with probability \\$\{0.30, 0.25, 0.24, 0.14, 0.07\}$ respectively.
\item 
A random respondent will answer the fifth question with grade $\{1, 2, 3, 4, 5\}$ with probability \\$\{0.38, 0.22, 0.22, 0.12, 0.06\}$ respectively.
\end{enumerate}

%grade
\begin{task}
\label{t:2quest3}
The respondent answered the first question with grade 5, the second and third questions with grade 3, the fourth question with grade 2, the fifth question with grade 3. What is the probability that the respondent has the trait $F_1$ with grade not less than 3? 
\end{task}
Corresponding R-code, which solves this Task:

\begin{verbatim}
>bnet.ev <- setEvidence(bnet, nodes 
= c("Q11","Q12","Q13","Q14","Q15"), states=c("5","3","3","2","3"))
> xq2 = querygrain(bnet.ev, nodes=c("F"))
> xq2
$F
F
0    1    2    3    4
0.05 0.36 0.33 0.11 0.14
\end{verbatim}
The results of the R-code can be interpreted as follows: From the last line in R environment one can see that on the basis of a priori data for values of trait $\{0,1,2,3,4\}$ we will have corresponding output probabilities $\{0.05, 0.36, 0.33, 0.11, 0.14\}$.
\begin{task}
\label{t:2quest4}
Assume that a respondent answered the first question with grade 5, the second and third questions with grade 3. What is the probability that the respondent will answer the fourth and fifth questions with grades not less than 4? 
\end{task}
Corresponding R-code, which solves this Task:
\begin{verbatim}
> bnet.ev <- setEvidence(bnet, nodes= c("Q11","Q12","Q13"),
states = c("5","3","3"))
> xq3 = querygrain(bnet.ev, nodes=c("Q14","Q15"))
> xq3
$Q14
Q14
1    2    3    4    5
0.29 0.26 0.24 0.15 0.07 
$Q15
Q15
1    2    3    4    5
0.34 0.25 0.23 0.19 0.06 
\end{verbatim}

The results of the R-code can be interpreted as follows: The respondent will answer the fourth question with grade not less than 4 with probability (15\%+7\%)=22\%. The respondent will answer the fifth question no worse than fourth with probability (19\%+6\%)=25\%.

\section{Conclusion}
We considered three different examples of psychological tests. The first test consisted of asking one question, the second of two questions and the third of five questions. After we set up all required statistical parameters and priors, we formulated three possible Tasks and offered their solutions in R environment. The solution includes the construction of a Bayesian network for each Task and computing posterior probabilities. 
We used the constructed Bayesian networks to develop principles for computing the overall grade of the given trait $F$ (for instance, the stress resistance). This overall grade tells us the level of possession of this trait $F$ by the given respondent. We demonstrated the potential of graphical probabilistic models of three simple examples. Finally, we showed the capabilities of Bayesian networks for qualitative analysis of the resulting solution.

Although we considered relative simple examples with just one trait, the offered technique and software can be used in cases with more traits. An example of a test case with more than one trait will be considered in a soon to be published paper but we also did not observe any restrictions or limitations in that work. The number of questions in each test can also be increased. The offered R-code only solves the described examples. However, this R-code can be modified for larger numbers of traits, questions and tests. 
\section{Software}
We use R programming language for realization of Bayesian networks due to its popularity among applied scientists/statisticians \cite{Bu:10,Kab:14}. The analogical work can be done in 
\begin{itemize}
\item 
MATLAB \cite{Mur:01};
\item
in one of the known software packages:
\begin{itemize}
\item 
% \cite{Gs:16}, 
GeNIe \& SMILE, \mbox{http://genie.sis.pitt.edu}
\item
% \cite{OB:11}, 
OpenBayes, \mbox{https://github.com/abyssknight/}
\item
% \cite{BA:04}; 
BANSY3, \mbox{http://www.dynamics.unam.edu/}\\ \mbox{DinamicaNoLineal3/bansy3.htm}
\end{itemize}
\item in one of the commercial products: AgenaRisk Bayesian network tool, Bayesian network application library, Bayesia, BNet.
%\item in one of the commercial products: AgenaRisk Bayesian network tool \cite{AR:16}, Bayesian network application library \cite{BAY:15}, Bayesia \cite{BNAL:00}, BNet \cite{BNet:14}.
\end{itemize}
\subsection{Reproducibility} To reproduce the presented results one can download the R-code from Dropbox\\
  \linkurl{https://www.dropbox.com/sh/t8cm12vv741a0h0/AABz_SwBEQ5mgKMyRAcl51mZa?dl=0}.
%\href{http://google.com}{or by clicking here}.

%
%

\section*{Acknowledgment}
This work was supported by the Institute of Mathematics and Mathematical Modeling CS MES Republic of Kazakhstan and by the Ministry of Education and Science of Kazakhstan, (grant number is 4085/GF4, 0115RK00640). 

Additionally, we would like to express our enormous gratitude to Prof. Maksat Kalimoldayev, 
Academic member of the National Academy of Sciences, the head of Institute of Information and Computational Technologies CS MES Republic of Kazakhstan 
for his organizational assistance, valuable comments and financial support.

\section*{Author biographies}
\begin{wrapfigure}{l}{25mm} 
\includegraphics[width=1in,height=1.2in,clip,keepaspectratio]{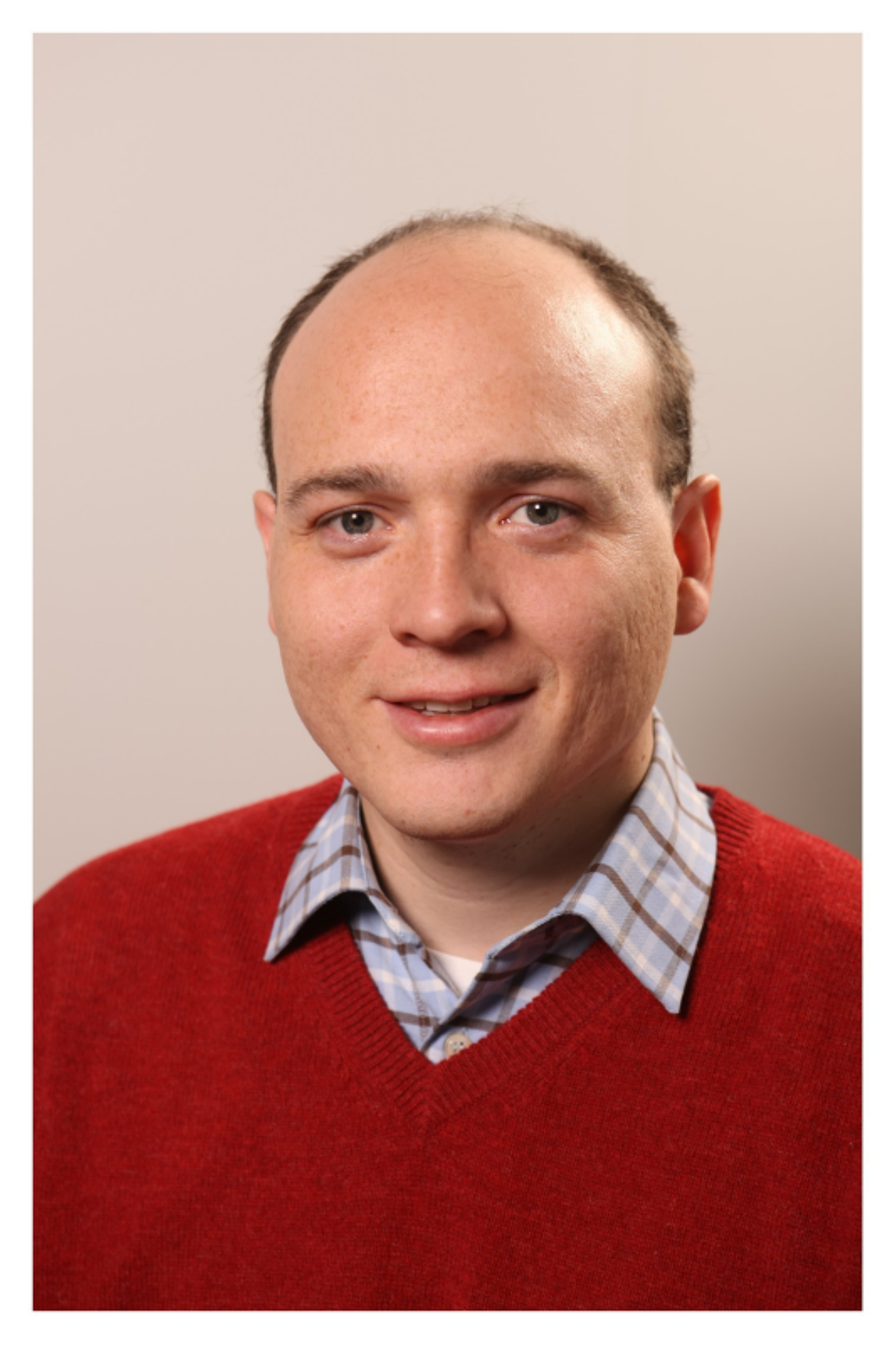}\end{wrapfigure}\par
{\bf Alexander Litvinenko} joined KAUST in 2013. He specializes in efficient numerical methods for stochastic PDEs, uncertainty quantification, multi-linear algebra and Bayesian update methods. Alexander earned B.Sc. (2000) and M.Sc. (2002) degrees in mathematics at Novosibirsk State University, and his PhD (2006) at Max-Planck-Institut in Leipzig. From 2007-2013 he was a Postdoctoral Research Fellow at the TU Braunschweig in Germany.\\
\linkurl{https://ecrc.kaust.edu.sa/Pages/Litvinenko.aspx}\\

\begin{wrapfigure}{l}{25mm} 
\includegraphics[width=1in,height=1.2in,clip,keepaspectratio]{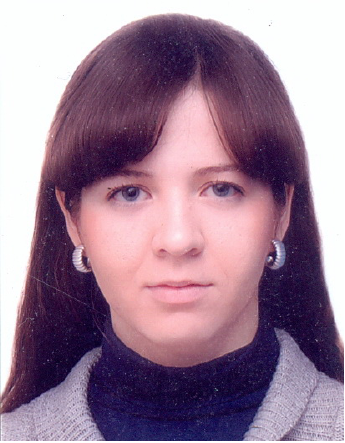}\end{wrapfigure}\par
{\bf Natalya Litvinenko} earned B.Sc. (2012) and M.Sc. (2014) in Mathematics at the Kazakh National University named after Al-Farabi, she also did an internship at Imperial College London (2014). Her research interest includes implementation of parallel algorithms on architectures NVIDIA CUDA, and Bayesian Networks. From 2015 till now she is taking part in the project ``Automated techniques for social-psychological diagnostics of military teams".\\

\begin{wrapfigure}{l}{25mm} 
\includegraphics[width=1in,height=1.2in,clip,keepaspectratio]{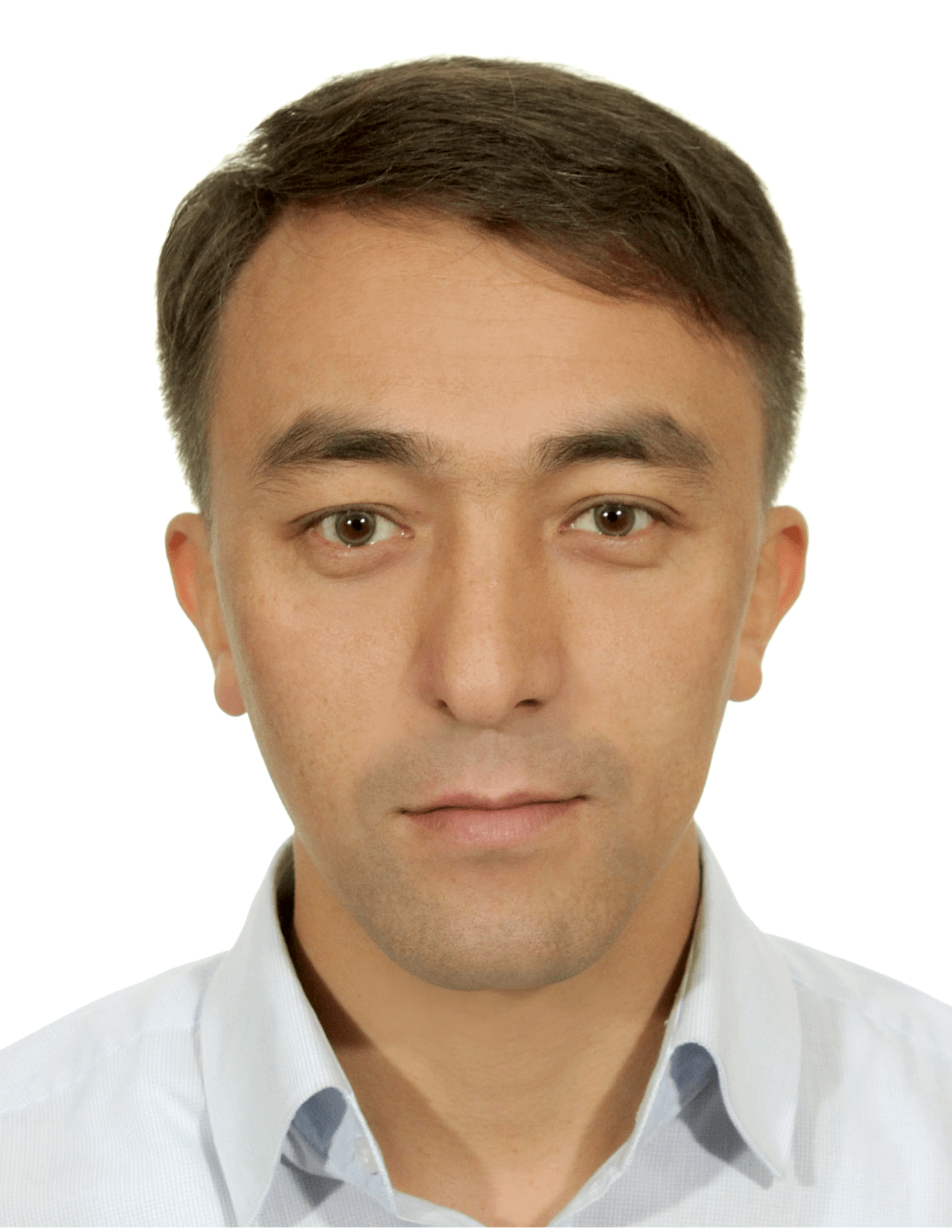}\end{wrapfigure}\par
{\bf Orken Mamyrbayev} earned B.Sc. (2001) and M.Sc. (2004) in Information systems at the Kazakh National Research Technical University named after K.I.Satpayev. His research interest includes digital signal processing, robotic systems, computer vision, automatic speech recognition. He earned his PhD (2014) in the group of Prof. Maksat Kalimoldayev. Currently he is deputy director in the Institute of Information and Computational Technologies CS MES RK.
\end{document}